\documentclass[aps,prb,twocolumn,preprintnumbers,amsmath,amssymb,superscriptaddress,showpacs,longbibliography]{revtex4-2}
\usepackage[latin9]{inputenc}
\usepackage{mathrsfs}
\usepackage{amsmath}
\usepackage{graphicx}
\usepackage{latexsym}
\makeatletter

\DeclareTextSymbolDefault{\textquotedbl}{T1}

\makeatother

\begin{document}
\title{Energy Dispersion, Superconductivity and Magnetic Fluctuations in Stacked Altermagnetism
Materials}
\author{Jun Chang}
\email{junchang@snnu.edu.cn}

\affiliation{School of Physics and Information Technology, Shaanxi Normal University,
Xi'an 710119, China}
\author{Hantao Lu}
\email{luht@lzu.edu.cn}

\affiliation{Key Laboratory of Quantum Theory and Applications of MoE, Lanzhou
University, Lanzhou 730000, China}
\affiliation{Lanzhou Center for Theoretical Physics, Key Laboratory of Theoretical
Physics of Gansu Province, Lanzhou University, Lanzhou 730000, China.}
\author{Jize Zhao}
\affiliation{School of Physical Science and Technology $\&$ Key Laboratory of
Quantum Theory and Applications of MoE, Lanzhou University, Lanzhou
730000, China.}
\affiliation{Lanzhou Center for Theoretical Physics, Key Laboratory of Theoretical
Physics of Gansu Province, Lanzhou University, Lanzhou 730000, China.}
\author{Hong-Gang Luo}
\affiliation{School of Physical Science and Technology $\&$ Key Laboratory of
Quantum Theory and Applications of MoE, Lanzhou University, Lanzhou
730000, China.}
\affiliation{Lanzhou Center for Theoretical Physics, Key Laboratory of Theoretical
Physics of Gansu Province, Lanzhou University, Lanzhou 730000, China.}
\author{Yang Ding}
\affiliation{Center for High-Pressure Science and Technology Advanced Research,
Beijing 100094, China}
\begin{abstract}
Recently, altermagnetism (AM) has emerged as a new category of magnetism,
alongside conventional antiferromagnetism (AFM) and ferromagnetism
(FM). In an AM, superconductivity (SC) is faced with a dilemma that
the spin-polarized bands, induced by the broken time reversal ($\mathcal{T}$)
symmetry, dominantly supports spin-triplet pairing. In contrast, AM
spin fluctuations routinely facilitate spin-singlet pairing as in
AFM. Consequently, unconventional SC is either absent or weak in AM
materials. Here, we propose that stacking 2D AM materials could resolve
this dilemma. Stacked 2D materials have yielded a variety of new electronic
properties by altering the symmetries inherent in the monolayer. In
a 2D anisotropic Hubbard model, we investigate the general energy
dispersions of both single-layer and stacked AM materials. We demonstrate
that AM sheet stacking can alter the original symmetries, consequently
affecting the energy dispersion. The interlayer magnetic coupling
enhances the low $\mathbf{q}$ magnetic fluctuations. $\mathcal{T}$
symmetry is restored in the AA stacking with an antiferromagnetic
interlayer coupling, and then both the energy dispersion and pairing
interaction are in favor of spin-singlet SC. The ferromagnetic interlayer
coupling in the AB stacking not only recovers $\mathcal{T}$ symmetry
but also supports spin-triplet pairing. It is further anticipated
that twisted bilayer AM sheets could exhibit additional novel electronic
properties, including topology, flat bands, and collective excitations.
Our work illustrates that stacking sheets of AM materials could open
up a unique research domain in exploring novel quantum phenomena and
offer a fertile ground for potential electronic applications. 
\end{abstract}
\date{\today }
\maketitle

\section{Introduction }

A recently identified category of magnetism, referred to as AM, has
been extensively studied due to its distinct properties and potential
applications \cite{Hayami2019,PhysRevX.12.031042,Hayami2020,Smejkal2020,mazin2021prediction,PhysRevB.102.014422,PhysRevLett.132.236701,PhysRevB.110.144412}.
It exhibits a zero net magnetization, akin to conventional AFM, and
spin-splitting energy dispersion similar to FM in the reciprocal momentum
space. Owing to their significant fundamental and technological significance,
extensive theoretical and experimental research has been conducted
on these captivating magnets. A variety of unique electromagnetic
phenomena have been theoretically predicted, including spin-dependent
band splitting, anomalous Hall and Kerr effects \cite{vsmejkal2022anomalous,PhysRevB.102.075112,PhysRevB.104.024401,samanta2020crystal,PhysRevB.99.184432},
as well as spin current and torque \cite{naka2019spin,ma2021multifunctional,PhysRevLett.126.127701,shao2021spin,PhysRevB.108.L180401,PhysRevLett.128.197202},
and substantial tunneling magnetoresistance effects \cite{PhysRevX.12.011028}.
Meanwhile, materials hypothesized to exhibit antiferromagnetic properties
have been the subject of experimental inquiries, such as $\mathrm{RuO_{2}}$
\cite{Berlijn2017,PhysRevX.12.031042,bose2022tilted,feng2022anomalous,Fedchenko2024},
$\mathrm{FeSb_{2}}$ \cite{PhysRevX.12.040002}, $\mathrm{MnF_{2}}$
\cite{PhysRevB.102.014422,PhysRevX.14.011019,Moreno2016}, $\mathrm{MnTe}$
\cite{Lee2024,Osumi_2024,krempasky2024}, $\mathrm{Mn_{5}Si_{3}}$
\cite{reichlova2021}, $CrSb$ \cite{Reimers2024} and $\mathrm{La_{2}CuO_{4}}$
\cite{PhysRevX.12.031042}.

\begin{figure}[th]
\includegraphics[width=1\columnwidth]{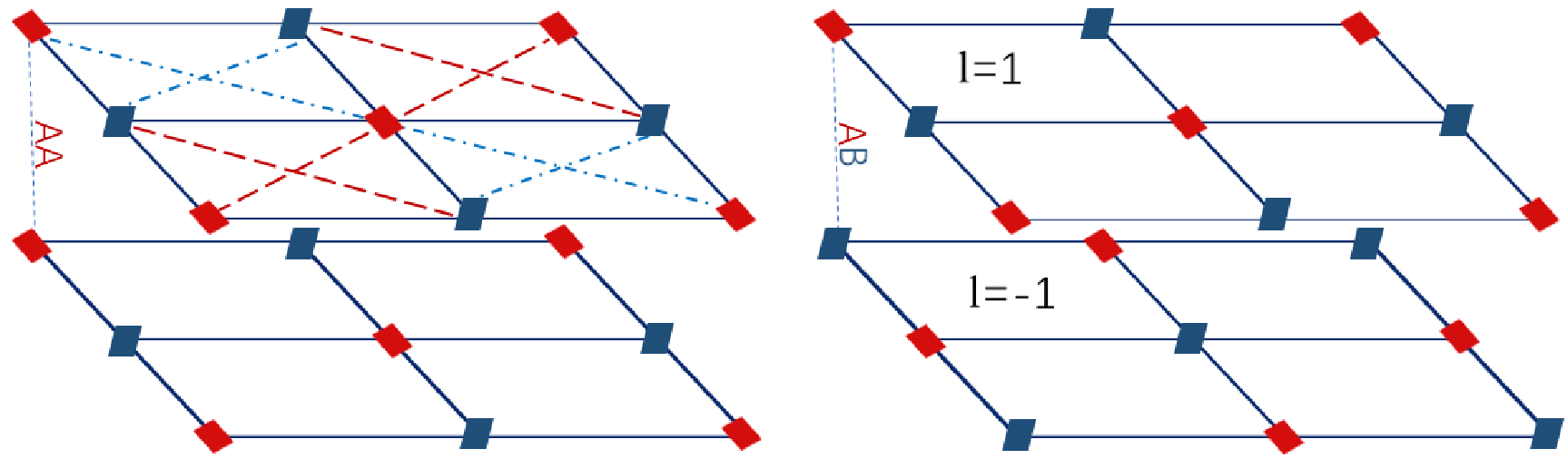} \caption{Schematic of stacking 2D altermagnetism sheets without a twist
angle. The lattice is divided into A (red) and B (blue) sublattices,
with distinct hopping integrals between the same-type sublattices.
The index $l=1$ and $-1$ signify the top and the bottom layers,
respectively. Within the same layer, the average spin polarizations
on A and B sublattices are antiparallel, with $S_{A}=-S_{B}$, whereas
across different layers, the spin polarizations may either maintain
this antiparallel relationship, $S_{A}(l=1)=-S_{A}(l=-1)$, or align
parallel, $S_{A}(l=1)=S_{A}(l=-1)$. The patterns of A-A and B-B stacking
are depicted in the left and right panels, respectively. \label{fig1}}
\end{figure}

The emergence of SC in AM has garnered immediate interest \cite{mazin2022,Zhu2023,Beenakker2023,PhysRevB.108.224421,Zhang2024,PhysRevLett.133.106601,PhysRevB.108.205410,PhysRevB.110.205120,chakraborty2024,Hong2024}.
A defining characteristic of AM is the broken translation or inversion
symmetry between the two magnetic sublattices besides the antiparallel
magnetization. Owing to the alternating spin polarization in real
space, the superconducting condensate in AM remains nonmagnetic, similar
to that in AFM. Unlike AFM, the AM structure, similar to FM, breaks
the $\mathcal{T}$ symmetry and eliminates the spin degeneracy of
the electronic energy bands, resulting in $\varepsilon(\mathbf{k},\uparrow)\neq\varepsilon(-\mathbf{k},\downarrow)$.
On one hand, near the Fermi energy levels, spin-polarized electrons
with antiparallel momenta tend to form spin-triplet Cooper pairs,
as is typical in FM. On the other hand, the superconducting pairing
interaction is frequently ascribed to magnetic fluctuations in both
AFM and FM, which support spin-singlet and spin-triplet SC, respectively.
Given that the AM order typically deviates only slightly from AFM,
the spin fluctuation spectra in AM closely resemble those in AFM \cite{mazin2022,Maier2023}.
Consequently, the SC induced by AM magnetic fluctuations is expected
to be predominantly characterized by spin-singlet pairing. This conflict
between the pairing interactions and the electronic structure could
potentially lead to the absence or weakness of SC in AM. Despite the
potential for magnetic fluctuations at low $\mathbf{q}$, arising
from scattering between different spin-polarized bands, to support
spin-triplet SC in AM, the typically weak spectral weight of these
fluctuations at low $\mathbf{q}$ results in substantially weaker
spin-triplet pairing \cite{Maier2023}.

In this study, we concentrate on the energy dispersion of 2D AM materials.
A general anisotropic Hubbard model, incorporating an on-site Coulomb
repulsion denoted by U is investigated. Utilizing the onsite mean
field approximation, we derive the energy dispersion. The unconventional
superconducting pairing is ascribed to the short-range magnetic interactions
among electrons. In a single layer, the energy band is spin split
due to the broken $\mathcal{T}$ symmetry by the AM order; thus spin-singlet
Cooper pairs with $\mathcal{T}$ symmetry are not favored. On the
other hand, AM spin fluctuations typically resemble those in AFM,
favoring spin-singlet pairing. We propose that stacking AM sheets in
some patterns can restore the broken $\mathcal{T}$ symmetry in AM
materials, thereby supporting the formation of spin-singlet Cooper
pairs. The interlayer magnetic coupling strengthens the low $\mathbf{q}$
magnetic fluctuations. Stacking sheets of 2D materials alters the
original symmetries of the individual layers~\cite{PhysRevLett.132.263402,giuli2025,delre2024}
and endows them with a variety of new electronic properties \cite{cao2018unconventional,PhysRevX.8.031088,PhysRevLett.99.256802,PhysRevB.81.161405,PhysRevB.82.121407,Bistritzer2011}.
Furthermore, when the layers are twisted relative to each other at
an angle, the resulting superlattice exhibits a significantly larger
periodicity than the individual layer, thereby possessing complex
and rich electronic properties, such as superconductivity, correlated
insulating states, and flat electronic bands. We anticipate that stacked
2D AM materials will pave new avenues for both fundamental research
and potential applications. The capability to modulate electronic
properties through twist angle variation and external field application
further constitutes a crucial platform for investigating novel quantum
phenomena in AM.

\section{Single layer AM model}

Initially, to investigate the properties of a single AM sheet, we
construct a 2D Hubbard model with the Hamiltonian

\begin{equation}
H=-\sum_{ij\sigma}\left(t_{ij}+\mu\delta_{ij}\right)c_{i\sigma}^{\dagger}c_{j\sigma}+U\sum_{i}n_{i\uparrow}n_{i\downarrow},\label{H}
\end{equation}
where $c_{i\sigma}^{\dagger}$ and $c_{i\sigma}$ are the electron
creation and annihilation operators with spin $\sigma$ on site $i$.
$n_{i\sigma}=c_{i\sigma}^{\dagger}c_{i\sigma}$ represents the corresponding
particle number operators. Furthermore, $t_{ij}$ denotes the single-particle
hopping integral between site $i$ and $j$. $U$ represents the on-site
Coulomb repulsion, while $\mu$ is the chemical potential.

For the interaction term of Eq. (\ref{H}), we initially ignore the
fluctuations in particle number and subsequently apply the mean field
decoupling \cite{Maier2023},

\begin{equation}
Un_{i\uparrow}n_{i\downarrow}\approx U\left(\left\langle n_{i\uparrow}\right\rangle n_{i\downarrow}+n_{i\uparrow}\left\langle n_{i\downarrow}\right\rangle -\left\langle n_{i\uparrow}\right\rangle \left\langle n_{i\downarrow}\right\rangle \right).\label{Hmf-1}
\end{equation}
Further, the average spin and particle number are defined, 
\begin{equation}
S_{i}\equiv\frac{1}{2}\left\langle n_{i\uparrow}-n_{i\downarrow}\right\rangle \text{ \& }N_{i}\equiv\left\langle n_{i\uparrow}+n_{i\downarrow}\right\rangle .\label{Hmf-2}
\end{equation}
The mean field Hamiltonian is written as 
\begin{equation}
H_{mf}=-\sum_{ij\sigma}\left(t_{ij}+\mu\delta_{ij}+\sigma US_{i}\delta_{ij}\right)c_{i\sigma}^{\dagger}c_{j\sigma},\label{Hmf}
\end{equation}
where $UN_{i}$ has been absorbed into the chemical potential, and
the constant term $U\left\langle n_{i\uparrow}\right\rangle \left\langle n_{i\downarrow}\right\rangle $
is neglected. To characterize AM, the lattice is partitioned into
A and B sublattices, featuring distinct intrasublattice (AA and BB)
hopping terms. By setting the chemical potential $\mu$ to zero, the
Fourier transformation of the mean field Hamiltonian into reciprocal
space yields 
\begin{equation}
H_{mf}=-\sum_{aa'\mathbf{k}\sigma}\varepsilon_{aa'}\mathbf{\left(\mathbf{k},\sigma\right)}c_{a\mathbf{k}\sigma}^{\dagger}c_{a'\mathbf{k}\sigma}\label{Hk}
\end{equation}
where $\mathbf{k}$ is the crystal momenta of a sublattice, $a$ and
$a'$ belong to sublattice $A$ or $B$, and the matrix elements of
the Hamiltonian are given by 
\begin{equation}
\varepsilon_{aa'}\left(\mathbf{k},\sigma\right)=\epsilon_{aa'}(\mathbf{k})+\sigma US_{a}\delta_{aa'}\label{Eks}
\end{equation}
where $\sigma=\pm1$ for spin up and down, respectively, and $\epsilon_{aa'}(\mathbf{k})$
represents the spin independent matrix elements, determined by the
lattice structure and the hopping integrals between sublattices $a$
and $a'$. The nature of Eq. (\ref{Eks}) is highly sensitive to
the hopping integrals $t_{ij}$, the mean spin values $S_{A}$ and
$S_{B}$ on the two sublattices. When there is no net magnetization
on either sublattice, or $S_{A}$=$S_{B}=0$, it is a paramagnetic
(PM) state. A Ferromagnetic state is realized when both sublattices
have the parallel alignment magnetization, or $S_{A}$=$S_{B}\neq0$.
The antiferromagnetic state emerges when there is translation or inversion
symmetry between the two antiparallel magnetic sublattices with $S_{A}$=
-$S_{B}\neq0$. Lastly, an altermagnetic state requires the breaking
of translation and inversion symmetry between the two antiparallel
magnetic sublattices. This can be achieved through anisotropic hopping
constants or a rotation symmetry between the two sublattices, along
with nonzero magnetizations. In AM materials, it is assumed that the
electron density is uniform with $N_{A}=N_{B}$, whereas the spin
polarization alternates between sublattices A and B with $S_{A}=-S_{B}$.
Then, the system energy dispersion is given by

\begin{equation}
\varepsilon^{\mp}(\mathbf{k},\sigma)=\varepsilon_{A+B}(\mathbf{k},\sigma)\mp\sqrt{\epsilon_{AB}^{2}(\mathbf{k})+\varepsilon_{A-B}^{2}(\mathbf{k},\sigma)}\label{Emfk}
\end{equation}
where 
\[
\varepsilon_{A\pm B}(\mathbf{k},\sigma)\equiv\frac{1}{2}\left[\varepsilon_{AA}(\mathbf{k},\sigma)\pm\varepsilon_{BB}(\mathbf{k},\sigma)\right]
\]
and 
\[
\epsilon_{A\pm B}(\mathbf{k})\equiv\frac{1}{2}\left[\epsilon_{AA}(\mathbf{k})\pm\epsilon_{BB}(\mathbf{k})\right].
\]
With the aid of Eq. (\ref{Eks}), $\varepsilon_{A+B}(\mathbf{k},\sigma)=\epsilon_{A+B}(\mathbf{k})$
is spin independent and $\varepsilon_{A-B}(\mathbf{k},\sigma)=\epsilon_{A-B}(\mathbf{k})-\sigma h$
is spin dependent, where $h=US_{A}$. Then the system energy dispersion
in Eq. (\ref{Emfk}) is rewritten as

\begin{equation}
\varepsilon^{\mp}(\mathbf{k},\sigma)=\epsilon_{A+B}(\mathbf{k})\mp\sqrt{\epsilon_{AB}^{2}(\mathbf{k})+\left[\epsilon_{A-B}(\mathbf{k})-\sigma h\right]^{2}}.\label{E1}
\end{equation}
Given that $\epsilon_{aa'}(\mathbf{k})$ exhibits inversion symmetry
in the momentum space, i.e. $\epsilon_{aa'}(\mathbf{k})=\epsilon_{aa'}(-\mathbf{k})$,
it follows that the spin-splitting bands $\varepsilon^{\mp}(\mathbf{k},\uparrow)\neq\varepsilon^{\mp}(-\mathbf{k},\downarrow)$
provided that the translation or inversion symmetry between the A
and B sublattice is broken, i.e. $\epsilon_{A-B}(\mathbf{k})\neq0$.
While $\epsilon_{A-B}(\mathbf{k})=0$ leads to spin degenerated AFM
state. In the absence of spin-orbit interaction, such energy dispersion
exclusively supports spin-triplet Cooper pairs with parallel spin
order parameter $\Delta_{\sigma\sigma}$, and no coupling exists between
the two order parameters $\Delta_{\uparrow\uparrow}$ and $\Delta_{\downarrow\downarrow}$.
The spin symmetry between up and down in AM sublattice ensures the
average order parameter unitary \cite{mazin2022}.

\begin{figure}[th]
\includegraphics[width=1\columnwidth]{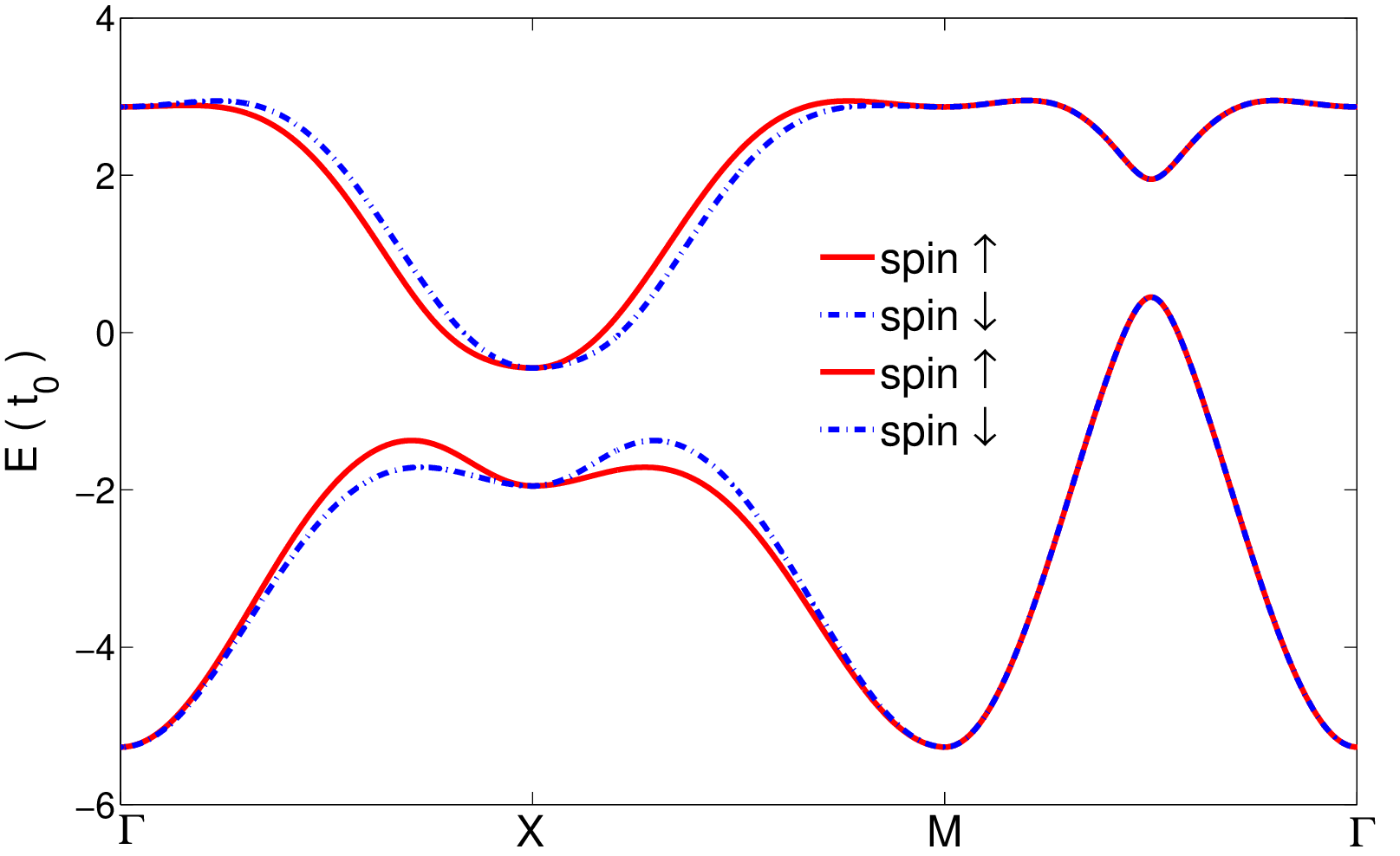} \caption{Metallic AM energy band structure on a square lattice in Eq. (\ref{E1}).
Following the parameters in Ref. \cite{Maier2023}, $t_{0}=1$
as the unit of energy, $t_{1}=0.4t_{0}$, $t_{2}=0.2t_{0}$ and $U=3.6t_{0}$.
Red and blue bands indicate the spin-$\uparrow$ and spin-$\downarrow$
components, respectively. The spin splitting is $\mathbf{k}$-dependent.
Along $M-\Gamma$, the spin-$\uparrow$ and spin-$\downarrow$ bands
are degenerate. The Fermi surfaces consist of hole pockets at $(\pm\pi/2,\pm\pi/2)$
and electron pockets at $(\pm\pi,0)$ and $(0,\pm\pi)$ in the reduced
Brillouin zone. \label{fig2}}
\end{figure}

Eq. (\ref{E1}) describes a general energy dispersion relation for
anisotropic AFM and AM materials. For AFM materials, there is typically a translation or inversion symmetry between the two sublattices with opposite spins. This symmetry implies that the spin-independent Hamiltonian matrix elements $\epsilon_{aa}(\mathbf{k})$ are identical for both sublattices, i.e., $\epsilon_{AA}(\mathbf{k}) = \epsilon_{BB}(\mathbf{k})$ or  $\epsilon_{A-B}(\mathbf{k}) = 0$ for all $\mathbf{k}$. Given this symmetry, the energy dispersions are spin-degenerate, $\varepsilon^{\mp}(\mathbf{k},\sigma)=\epsilon_{A+B}(\mathbf{k})\mp\sqrt{\epsilon_{AB}^{2}(\mathbf{k})+h^{2}}$. On the other hand, if the translation or inversion symmetry between the A and B sublattices is broken, then there exists $\mathbf{k}$ such that  $\epsilon_{A-B}(\mathbf{k})\neq0$. In this case, the energy dispersions become spin-dependent, which is characteristic of altermagnetism. For example, if the translation or inversion transformation of the A sublattices is followed by a rotation operation and spin reversal operation $\bar 1$ to reach the B sublattices \cite{PhysRevX.14.031038}, then, the translation or inversion symmetry between the A and B sublattices can be broken. The broken symmetry can lead to differences in the Hamiltonian matrix elements between the two sublattices, specifically, $\epsilon_{A-B}(\mathbf{k})\neq0$. This results in spin-dependent energy dispersions. In Fig. \ref{fig1}, the translation transformation of the A sublattices is followed by a $\pi/2$ rotation about the axis perpendicular to the planes and spin reversal $\bar 1$ to reach the B sublattices. Once the specific geometric structure is defined, the
specific type of spin splitting can be determined. Considering the
square lattice with two sublattices as an example, the matrix elements
of the Hamiltonian are given by \cite{Maier2023} 

\begin{equation}
\epsilon_{AA}\left(\mathbf{k}\right)=-2t_{1}\cos2k_{x}-2t_{2}\cos2k_{y},\label{Eks-1}
\end{equation}
\begin{equation}
\epsilon_{BB}\left(\mathbf{k}\right)=-2t_{2}\cos2k_{x}-2t_{1}\cos2k_{y},\label{Eks-1-1}
\end{equation}
\begin{equation}
\epsilon_{AB}\left(\mathbf{k}\right)=-2t_{0}\left(\cos k_{x}+\cos k_{y}\right)\label{Eks-1-1-1}
\end{equation}
where $t_{0}$ represents the nearest sublattice A-B hopping amplitude,
$t_{1}$ and $t_{2}$ are the intrasublattice hopping constants that
break the sublattice $C4$ symmetry. Adopting the parameters from
Ref. \cite{Maier2023}, with $t_{0}=1$ as the unit of
energy, $t_{1}=0.4t_{0}$, $t_{2}=0.2t_{0}$ and $U=3.6t_{0}$. The
corresponding energy band structure, as described in Eq. (\ref{E1}),
is depicted in Fig. \ref{fig2}. The four energy bands result from
the sublattices and spins. The spin up and down bands are split, depending
on the crystal momentum when $t_{1}\neq t_{2}$. 

In the study of energy dispersion, we focus on the Hubbard model,
while, in the study of superconductivity, we turn to the low-energy
effective model of the Hubbard model, namely the $t-J$ model. The
Heisenberg exchange interaction couples electrons at neighboring sites
\begin{equation}
H_{int}=\sum_{ij}J_{ij}\mathbf{s}_{i}\cdot\mathbf{s}_{j}\label{Hint}
\end{equation}
where the spin operator $\mathbf{s}_{i}=c_{i\sigma}^{\dagger}\mathbf{\left[\boldsymbol{\boldsymbol{\sigma}}\right]_{\sigma\sigma'}}c_{i\sigma'}/2$
with Pauli matrix $\boldsymbol{\boldsymbol{\sigma}}=(\sigma_{x},\sigma_{y},\sigma_{z})$.
The pairing order parameters $\Delta_{\sigma\sigma'}$ could stem
from the magnetic interaction in Eq. (\ref{Hint}) using a mean field
method. From the real space point of view, the nearest-neighbor magnetic
exchange interaction $J_{AB}$ in the AM state typically favors spin-singlet
pairing and the next-nearest-neighbor $J_{AA}$ and $J_{BB}$ support
spin-triplet pairing. Nevertheless, the next-nearest-neighbor exchange
interactions are typically much weaker than the nearest-neighbor couplings.
Therefore, the spin-singlet pairing should be dominant in terms of
magnetic interaction. On the other side, the long-range magnetic order
and spin fluctuations near $\mathbf{q}\approx0$ in the AM phase are also
possible \cite{mazin2022}. Such magnetic fluctuations could support
triplet pairing. However, the spectral weight of the spin fluctuations
near $\mathbf{q}\approx0$ in AM is typically quite small, as the
AM order often slightly deviates from the AFM order \cite{Maier2023}.
Consequently, the spin fluctuation dominantly mediates spin-singlet
pairing. The contradiction between the electronic structure, favoring
spin-triplet pairing, and the pairing interaction, supporting spin-singlet
pairing, results in the absence of or weak SC in the AM phase. We
tend to alleviate this predicament by stacking AM sheets to alter
the electronic structures and spin fluctuation spectra.

\section{Stacked layer model }

Recently, atomically stacking quasi-2D materials results in various
superstructures by altering the original geometrical symmetries inherent
in individual layers \cite{cao2018unconventional,PhysRevX.8.031088,PhysRevLett.99.256802,PhysRevB.81.161405,PhysRevB.82.121407,Bistritzer2011}.
The artificial superstructure induces a variety of unique electronic
properties, including superconductivity, magnetism, topology, correlated
insulating states, and flat electronic bands. From a fundamental perspective,
the complexity and richness of the electronic properties in stacked
2D materials present significant opportunities to study emergent quantum
phenomena. From an application standpoint, stacked 2D materials offer
a unique potential for the development of low-power and high-speed
electronic devices. Specifically, when two sheets of a 2D material
are twisted at the \textquotedbl magic angle\textquotedbl , a moire
superlattice emerges due to the interference between the two layers.
Here, we consider two individual AM layers, designated by $l=\pm1$.
Initially, we disregard the interlayer coupling. When the local magnetic
moments on sublattice A are aligned, $S_{A}(l=1)=S_{A}(l=-1)$, the
energy dispersions mirror those in Eq. (\ref{E1}) for a single layer.
Conversely, if the local magnetic moments on sublattice A are in opposition,
$S_{A}(l=1)=-S_{A}(l=-1)$, then the dispersion is given by

\begin{equation}
\varepsilon^{\mp}(\mathbf{k},\sigma,l)=\epsilon_{A+B}(\mathbf{k})\mp\sqrt{\epsilon_{AB}^{2}(\mathbf{k})+\left[\epsilon_{A-B}(\mathbf{k})-l\sigma h\right]^{2}}.\label{E1-1}
\end{equation}
It is evident that the energy dispersion of the two sheets with opposite
spin orientations is identical, i.e. $\varepsilon^{\mp}(\mathbf{k},\uparrow,l=1)=\varepsilon^{\mp}(\mathbf{k},\downarrow,l=-1)$.
It is anticipated that the layer coupling could render the system
dispersion spin independent.

To unite two layers together, the layer coupling is introduced, such
as van der Waals interaction or chemical bonding. It is widely recognized
that the energy dispersion of stacked materials is contingent upon
the stacking patterns. For simplicity, we restrict our analysis to
zero twist angle between sheets and the effective interlayer hopping
$t_{\perp}(il,jl')c_{i\sigma l}^{\dagger}c_{j\sigma l'}$ is limited
to the nearest-neighbors, with $t_{\perp}(il,il')=t_{\perp}$. 

\subsection{AA stacking}

The nature of the interlayer magnetic coupling between the stacked
sheets, whether ferromagnetic or antiferromagnetic, is dictated by
a combination of the magnetic ions and their surrounding ligands,
which is encapsulated by the Goodenough-Kanamori-Anderson rule \cite{PhysRev.79.350,PhysRev.100.564,Kanamori1957a,Kanamori1957b}.
For the AA stacked AM sheets with $S_{A}(l=1)=-S_{A}(l=-1)$, the
mean field Hamiltonian of this system can be diagonalized exactly,
and the resulting energy dispersion is characterized by 
\begin{equation}
\varepsilon_{\mp}^{\mp}(\mathbf{k},\sigma)=\epsilon_{A+B}(\mathbf{k})\mp\sqrt{\epsilon_{AB}^{2}(\mathbf{k})+\xi_{\mp}^{2}(\mathbf{k})}\label{E2}
\end{equation}
with 
\begin{eqnarray}
\xi_{\mp}^{2}(\mathbf{k})= &  & \epsilon_{A-B}^{2}(\mathbf{k})+t_{\bot}^{2}+h^{2}\mp\nonumber \\
 &  & 2\sqrt{\epsilon_{A-B}^{2}(\mathbf{k})\left(t_{\bot}^{2}+h^{2}\right)+\epsilon_{AB}^{2}(\mathbf{k})t_{\bot}^{2}}.\label{E2-1}
\end{eqnarray}

It is evident that the dispersion is spin independent, with $\varepsilon_{\mp}^{\mp}(\mathbf{k},\uparrow)=\varepsilon_{\mp}^{\mp}(\mathbf{-k},\downarrow)$.
The interlayer coupling restores the broken $\mathscr{\mathcal{T}}$
symmetry present in the monolayer. This electronic structure, resulting
from interlayer coupling, could host spin-singlet Cooper pairs, in
contrast to the single AM layer, which only supports spin-triplet
SC. Furthermore, the stacking pattern modifies the spin fluctuations.
The nearest neighbor magnetic interaction between two layers is described
by

\begin{equation}
H_{int}=\sum_{i,l\neq l'}J_{ill'}\mathbf{s}_{il}\cdot\mathbf{s}_{il'}\label{Hint-1}
\end{equation}
where the layer index $l,l'=\pm1$. The interlayer spin coupling within
the same uinit cell is expected to enhance the low $\mathbf{q}$
spin fluctuations. The susceptibility $\chi_{0}$ can be experessed
by \cite{Chang2007,PhysRevB.110.144412,Maier2023} 
\begin{equation}
\chi_{0\sigma\sigma'}(\mathbf{q},\omega)=\sum_{\mathbf{k}}\frac{A_{\mathbf{k},\mathbf{q}}\left[f(\varepsilon_{\mp}^{\mp}(\mathbf{k},\sigma))-f(\varepsilon_{\mp}^{\mp}(\mathbf{k+q},\sigma')\right]}{\left[\omega+i\eta-\varepsilon_{\mp}^{\mp}(\mathbf{k},\sigma)+\varepsilon_{\mp}^{\mp}(\mathbf{k+q},\sigma')\right]}\label{x}
\end{equation}
where $A_{\mathbf{k},\mathbf{q}}$ is the coherence factor and $f$
is the Fermi distribution function. Due to the energy band degeneracy
$\varepsilon_{\mp}^{\mp}(\mathbf{k},\uparrow)=\varepsilon_{\mp}^{\mp}(\mathbf{k},\downarrow),$
the $\chi_{0}$ at low $\mathbf{q\approx0}$ spin fluctuations
is dominantly determined by the magnetic scattering processes between
the degenerate bands. The magnetic interaction in stacking materials
can be decomposed into intralayer and interlayer pairing under mean
field method \cite{PhysRevLett.132.146002}. The AFM interlayer magnetic
fluctuations are capable of mediating $s$-wave spin-singlet pairing.
Thus, in AA stacking with antiferromagnetic coupling between two layers,
both electronic structure and spin fluctuations support spin-singlet
SC.

\begin{figure}[th]
\includegraphics[width=1\columnwidth]{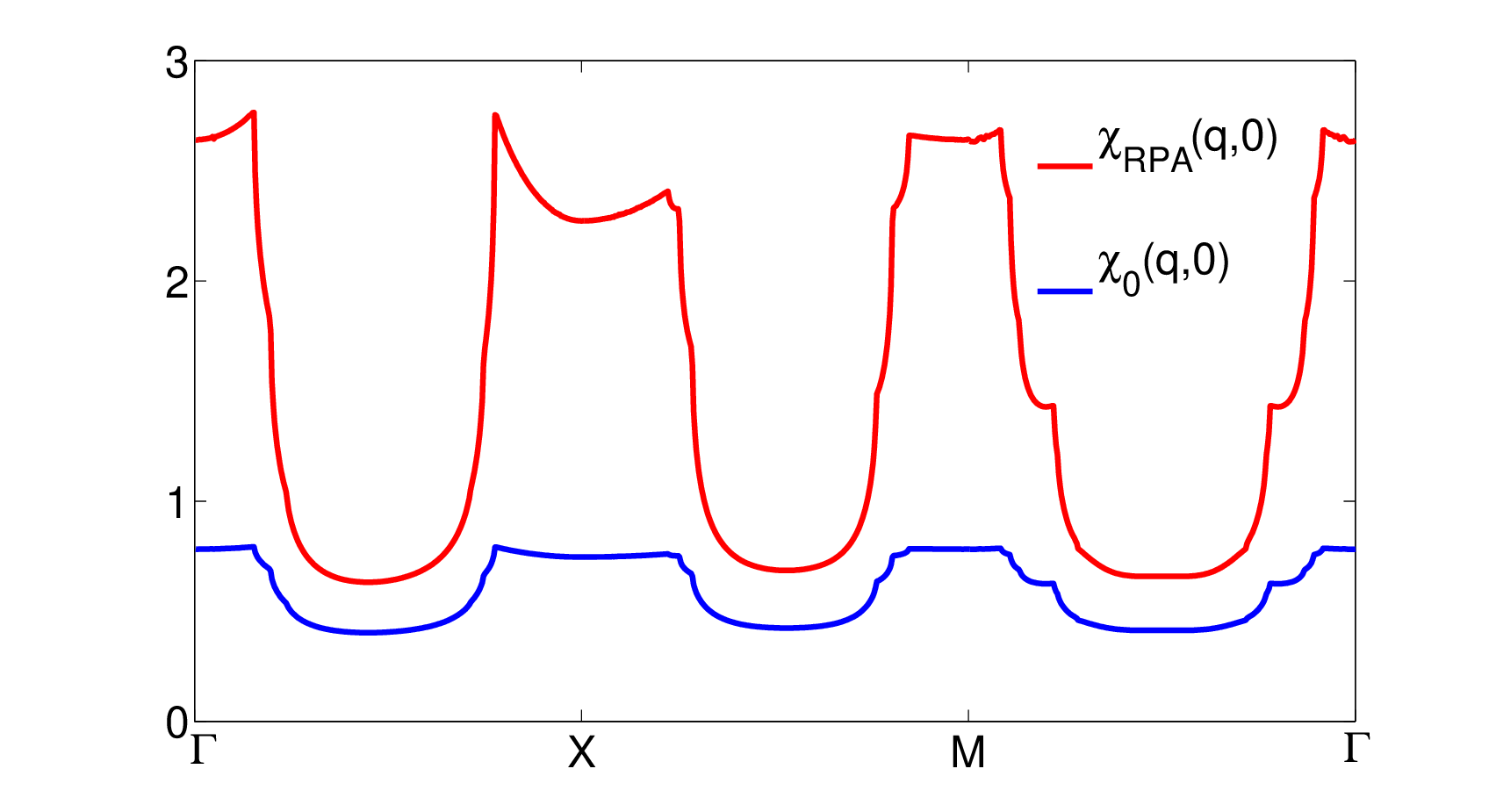} \caption{Bare and RPA magnetic susceptibilities $\chi_{0}(\mathbf{q},0)$ and $\chi_{RPA}(\mathbf{q},0)$ in the AA stacked AM sheets with $S_{A}(l=1)=-S_{A}(l=-1)$.
The temperatue $T =0.001t_0$, $t_{\perp} =0.2t_0$, and the other parameters are the same as those in Fig. \ref{fig2}. The RPA correction significantly strengthens the spin fluctuations $\chi_{RPA}(\mathbf{q},0)$ around the momenta $(0,0)$, $(\pi,0)$ and $(\pi,\pi)$. \label{fig3}}
\end{figure}

Using the same parameters as those for the square lattice with two sublattices in Fig. \ref{fig2}, we investigate the bare and random phase approximation (RPA) magnetic susceptibility of the AA stacked AM sheets with $S_{A}(l=1)=-S_{A}(l=-1)$. As depicted in Fig. \ref{fig3}, at the wave vector $\mathbf{Q}=(\pi,\pi)$, the bare susceptibility $\chi_{0}(\mathbf{Q},\omega=0)=\chi_{0}(\mathbf{q}=0,\omega=0)$ is proportional to the density of states, which is a consequence of the equality $\varepsilon_{\mp}^{\mp}(\mathbf{k})=\varepsilon_{\mp}^{\mp}(\mathbf{k+Q})$. The multiorbital RPA susceptibility
matrix can be written as \cite{PhysRevB.110.144412}
\begin{equation}
\left[\chi_{RPA}(\mathbf{q},iq_n)\right]^{\mu_1,\mu_2}_{\mu_3,\mu_4}=\left[\chi_{0}(\mathbf{q},iq_n)(1-U\chi_{0}(\mathbf{q},iq_n))^{-1}\right]^{\mu_1,\mu_2}_{\mu_3,\mu_4} \label{rpax}
\end{equation}
where $\mu_i$ is the orbital index and $iq_n$ is a bosonic Matsubara frequency, $\left[U\right]^{\mu_1,\mu_2}_{\mu_3,\mu_4}= U$
for $\mu_1=\mu_2=\mu_3=\mu_4$. The RPA correction of the onsite Hubbard interaction 
primarily amplifies the amplitude of spin fluctuations while subtly modifying the momentum structure. Furthermore,  in Fig. \ref{fig3}, the RPA correction significantly boosts the spin fluctuations $\chi_{RPA}(\mathbf{q},0)$ around the momenta $(0,0)$, $(\pi,0)$ and $(\pi,\pi)$. The enhancement is expected to favor interlayer SC pairing.

In the AA stacking, if the interlayer coupling is ferromagnetic or
$S_{A}(l=1)=S_{A}(l=-1)$, the energy dispersion becomes spin-dependent
and is given by 
\begin{equation}
\varepsilon_{\mp}^{\mp}(\mathbf{k},\sigma)=\epsilon_{A+B}(\mathbf{k})\mp t_{\bot}\mp\sqrt{\epsilon_{AB}^{2}(\mathbf{k})+\left[\epsilon_{A-B}(\mathbf{k})-\sigma h\right]^{2}}.\label{E2-2}
\end{equation}
The interlayer spin coupling significantly enhances the low $\mathbf{q}$
spin fluctuations by interband processes around energy $2t_{\bot}$
due to $\varepsilon_{\mp}^{-}(\mathbf{k},\sigma)=\varepsilon_{\mp}^{+}(\mathbf{k},\sigma)+2t_{\bot}$.
In contrast to the antiferromagnetic interlayer coupling, the ferromagnetic
coupling between layers favors s-wave spin-triplet pairing. Consequently,
the electronic structure and spin fluctuations both support spin-triplet
SC in AA stacking with ferromagnetic interlayer coupling. This distinction
underscores the pivotal role of interlayer magnetic coupling in determining
the type of superconducting pairing that can emerge in stacked systems.

Applying the decomposition of the Pauli matrix product into spin triplet
and singlet configurations, 
\begin{eqnarray}
\sigma_{\alpha\beta}\cdot\sigma_{\gamma\delta}=\frac{1}{2}\left(\delta_{\alpha\beta}\delta_{\gamma\delta}+\delta_{\alpha\delta}\delta_{\beta\gamma}\right)\nonumber \\
-\frac{3}{2}\left(\delta_{\alpha\beta}\delta_{\gamma\delta}-\delta_{\alpha\delta}\delta_{\beta\gamma}\right),\label{PTS}
\end{eqnarray}
the nearest neighbor magnetic interaction between two layers in spin-triplet
channel could be written as \cite{Chang2017}
\begin{equation}
J_{ll'}\mathbf{s}_{l}\cdot\mathbf{s}_{l'}\rightarrow\frac{J_{ll'}}{4}\left[n_{l\uparrow}n_{l'\uparrow}+n_{l'\downarrow}n_{l'\downarrow}\right]\label{Hint-1-1}
\end{equation}
where only the same spin interaction terms are kept, as the spin-split
bands in AM can only support Cooper pairs with the order parameter
$\Delta_{\sigma\sigma}$; see Ref. \cite{mazin2022}. The $s$-wave
spin-triplet parameter order in real space is $\Delta_{\sigma\sigma}(i)\sim\left\langle c_{il\sigma}c_{il'\sigma}\right\rangle |_{l\neq l'}$
for both sublattices, considering $\mathcal{T}$ symmetry broken by
the AM order, and after the Fourier transformation to moment space,
the pairing amplitudes are given by 
\begin{equation}
\Delta_{\sigma\sigma}^{s}=\frac{J_{ll'}}{4N}\sum_{\mathbf{k}}\left\langle c_{\mathbf{k}\sigma}c_{\mathbf{-k}\sigma}\right\rangle \left(\gamma^{s+is}\right)^{*}\label{sgap}
\end{equation}
where the ferromagnetic coupling $J_{ll'}<0$, and the $s$-wave gap
function factor $\gamma^{s+is}=\left(1+i\right)/\sqrt{2}$, indicating
broken $\mathcal{T}$ symmetry. Since SC requires a metallic altermagnetic
state, we assume the hole doping away from the Mott insulator is slight
and very close to half filling per sublattice. Then in the energy
dispersion Eq. (\ref{E2-2}), only the four lower $\varepsilon_{-}^{\mp}(\mathbf{k},\sigma)$
bands are occupied. Considering the slight doping, only the top $\varepsilon_{-}^{+}(\mathbf{k},\sigma)$
bands cross the Fermi energy level, and $\epsilon_{AB}(\mathbf{k})$
or the hopping between A and B sublattice is strongly suppressed by
the large Hubbard $U.$ Then 
\[
\varepsilon_{-}^{+}(\mathbf{k},\sigma)\approx\epsilon_{A+B}(\mathbf{k})+\sigma\epsilon_{A-B}(\mathbf{k})+t_{\bot}-h.
\]
Solving the SC gap Eq. (\ref{sgap}) self-consistently, the BCS-like
gap equation can be formulated as
\begin{eqnarray}
-\int\frac{d^{2}k}{4\pi^{2}}\frac{J_{ll'}}{4E^{s}(\mathbf{k},\sigma)}\tanh\frac{\beta E^{s}(\mathbf{k},\sigma)}{2}=1.\label{egap}
\end{eqnarray}
with the Bogoliubov quasiparticle energy spectra
\[
E^{s}(\mathbf{k},\sigma)=\sqrt{\left[\varepsilon_{-}^{+}(\mathbf{k},\sigma)\right]^{2}+\left[\Delta_{\sigma\sigma}^{s}\right]^{2}}.
\]
According to BCS theory, as long as $J_{ll'}<0$ or ferromagnetic
coupling, there exist a finite SC gap or order parameter to the gap
equation. 

The BCS-like gap equation for $S_{A}(l=1)=-S_{A}(l=-1)$ is analogous
to Eq. (\ref{egap}). The distinction lies in the emergence of SC
in spin-singlet channel, where the $s$-wave gap function factor $\gamma^{s}=1$,
exhibiting $\mathcal{T}$ symmetry, and the interaction term $J_{ll'}/4$
in the spin-triplet channel is replaced by $-3J_{ll'}/4$ in spin-singlet
channel gap equation according to the decomposition Eq. (\ref{PTS}).
For $J_{ll'}>0$ in antiferromagetic interlayer coupling, there exists
a finite SC order parameter solution to the gap equation.

\subsection{AB stacking}

For comparison, the dispersion in the AB stacking, characterized by
$S_{A}(l=1)=-S_{B}(l=-1)$, is spin-dependent, similar to that in
the single layer. The spin dependent dispersion is 
\begin{equation}
\varepsilon_{\mp}^{\mp}(\mathbf{k},\sigma)=\epsilon_{A+B}(\mathbf{k})\mp\sqrt{\left[\epsilon_{AB}(\mathbf{k})\mp t_{\bot}\right]^{2}+\left[\epsilon_{A-B}(\mathbf{k})-\sigma h\right]^{2}}.\label{E2-2-1}
\end{equation}

Though the antiferromagnetic spin coupling is capable of facilitating
spin-singlet pairing the energy dispersion $\varepsilon_{\mp}^{\mp}(\mathbf{k},\uparrow)\neq\varepsilon_{\mp}^{\mp}(\mathbf{-k},\downarrow)$
is unfavourable for the spin-singlet pairing formation, as in single-layer
AM.

Finally, the energy dispersion relation for the AB stacking configuration
with $S_{A}(l=1)=S_{B}(l=-1)$ is given by 
\begin{equation}
\varepsilon_{\mp}^{\mp}(\mathbf{k},\sigma)=\epsilon_{A+B}(\mathbf{k})\mp\sqrt{\epsilon_{A-B}^{2}(\mathbf{k})+\xi_{\mp}^{2}(\mathbf{k})}\label{E2-3}
\end{equation}
with 
\begin{eqnarray}
\xi_{\mp}^{2}(\mathbf{k})= &  & \epsilon_{AB}^{2}(\mathbf{k})+t_{\bot}^{2}+h^{2}\mp\nonumber \\
 &  & 2\sqrt{\left[\epsilon_{A-B}^{2}(\mathbf{k})+t_{\bot}^{2}\right]h^{2}+\epsilon_{AB}^{2}(\mathbf{k})t_{\bot}^{2}}.\label{E2-1-1}
\end{eqnarray}
The stacking restores $\mathcal{T}$ symmetry, thereby resulting in
$\varepsilon_{\mp}^{\mp}(\mathbf{k},\uparrow)=\varepsilon_{\mp}^{\mp}(\mathbf{-k},\downarrow)$
and the energy dispersion that is independent of spin. The electronic
structure is capable of supporting both spin singlet and triplet Cooper
pairing between layers. However, the ferromagnetic interlayer spin
fluctuations selectively enhance spin-triplet pairing between layers.

\section{Discussion and Conclusion}

In our model, we have ignored the spin-orbit coupling (SOC), electron-phonon
interaction, long-range interlayer coupling and twist angle between
the layers. Although omitting SOC is a simplification often made in
both the Hubbard and $t-J$ models the SOC breaks parity symmetry
and could induce a mixture of spin singlet and triplet pairing and
reconstruction of the Fermi surface. Electron-phonon interaction could
be conducive to the SC pairing. Considering the experimental realization of
the 2D single-layer and stacked AM models, a direct method is to exfoliate
single or double layers from these crystals and artificially assemble
the sheets. Presently, more than a dozen AM candidates have been suggested 
\cite{PhysRevX.12.031042}. The crystals with planar spin-momentum
locking, such as $La_{2}CuO_{4},FeSb_{2},KRu_{4}O_{8},RuO_{2},MnO_{2}$
and $MnF_{2}$, are the optimal candidates for exfoliating and assembling
operations.

To summarize, our study has investigated the energy dispersion in
both single and stacked layer AM materials. Using a 2D anisotropic
Hubbard model, we have analytically derived the general energy dispersions
using the mean field method. In the single AM sheet, due to $\mathscr{\mathcal{T}}$
symmetry broken, the energy dispersion is spin-polarized and only
spin-triplet SC is favored. In contrast, the spin fluctuations typically
favor spin-singlet pairing, as the AM order often slightly deviates
from that in AFM. The discord between electronic structure and pairing
interaction hinders SC formation in AM. Thus, we demonstrate that
stacking AM sheet could change the original symmetries present in
the monolayer, and interlayer magnetic coupling can enhance the low
$\mathbf{q}$ spin fluctuations. The stacking could make both the
electronic structure and spin fluctuations favor the formations of
SC including spin singlet and triplet pairings, depending on stacking
patterns. In the twisted bilayer AM layers, the unit cell expands,
complicating the interlayer coupling which becomes spatially dependent
rather than uniform. We further propose that twisted bilayer AM sheets
could give rise to additional novel electronic properties, such as
topology, flat bands, and collective excitations. Our work indicates
that stacking sheets of AM materials could provide a unique platform
for exploring new quantum phenomena. Modifying the electronic properties
by turning the twist angle between AM Sheets could lead to the development
of novel electronic devices.

\textit{Acknowledgments}.$-$ We are thankful to Yunhua Wang for fruitful
discussions. This work is supported by the National Natural Science
Foundation of China (Grants No.12174168, No.12247101, No.12274187,
No.11874188, No.12047501, No.11874075), Science Challenge Project
No. U1930401, and National Key Research and Development Program of
China 2018YFA0305703.

\end{document}